\documentclass[a4paper]{article}
\usepackage{INTERSPEECH2021}
\usepackage{amsmath,graphicx}
\usepackage{multirow}
\usepackage{graphicx}   
\usepackage{makecell}
\usepackage{subfigure}
\usepackage[para]{footmisc}
\usepackage{dcolumn}
\usepackage{color}
\usepackage{svg}
\svgsetup{
    inkscapepath=i/svg-inkscape/
}
\svgpath{{svg/}}

\newcommand{\tl}[1]{\multicolumn{1}{l|}{#1}} 

\title{ Multilingual Approach to Joint Speech and Accent Recognition with DNN-HMM Framework}
\name{Yizhou Peng$^{1*}$, Jicheng Zhang$^{1*}$, Haobo Zhang$^{1*}$, Haihua Xu$^2$, Hao Huang$^{1}$, Eng Siong Chng$^{2}$}
\address{$^1$School of Information Science and Engineering, Xinjiang University, Urumqi, China\\
  $^2$School of Computer Science and Engineering, Nanyang Technological University, Singapore}
\email{}
\begin{document}
\ninept
\maketitle

\newcommand\blfootnote[1]{%
\begingroup
\renewcommand\thefootnote{}\footnote{#1}%
\addtocounter{footnote}{-1}%
\endgroup
}


%
\begin{abstract}
Human can recognize speech, as well as the peculiar accent of the speech simultaneously. However, present state-of-the-art ASR system can rarely do that. In this paper, we propose a multilingual approach to recognizing English speech, and related accent that speaker conveys using DNN-HMM framework. Specifically, we assume different accents of English as different languages. We then merge them together and train a multilingual ASR system. 
During decoding, we conduct two experiments. 
One is a monolingual ASR-based decoding, with the accent information  embedded at phone level, realizing word-based accent recognition (AR), and the other is  a multilingual ASR-based decoding,
realizing an approximated utterance-based AR.
Experimental results on an 8-accent English speech recognition show both methods can yield WERs close to the conventional ASR systems that completely ignore the accent, as well as desired AR  accuracy. Besides, we conduct extensive analysis for the proposed method, such as transfer learning with out-domain data exploitation, cross-accent recognition confusion, as well as characteristics of accented-word.
%
\end{abstract}
%
\noindent\textbf{Index Terms}:
speech recognition, accent recognition, TDNNf, multilingual, multi-task learning

\section{Introduction}\label{sec:intro}
Automatic speech recognition~(ASR)  has been significantly improved thanks to the deep neural network
techniques~\cite{GOOGLE-ICASSP-2018-SRS2S,D.Povey-Interspeech-2018-E2E,FACEBOOK-2019-ASRU-hybrid, Karita-2019-interspeech}.
However, there still remains challenges in speech recognition area. For instance, state-of-the-art ASR performance would be severely degraded  when
it recognizes noisy speech~\cite{T.T-2018-TASLP-NRSR,D.Yu-2013-ICASSP-NRSR}. 
Another limitation is that most present  ASR systems are only capable of recognizing monolingual speech. However, as globalization trends are
deepened, multilingual ASR systems ~\cite{FACEBOOK-2020-Arxiv-MLASR,J.Cho-2018-SLT-MLASR,S.W-ASRU-LID-2017} are greatly required.
\blfootnote{~~Students ($^*$) have joined SCSE MICL Lab, NTU, Singapore as exchange students.
This work is supported by the National Key R\&D Program of China (2017YFB1402101), Natural Science Foundation of China (61663044, 61761041), Hao  Huang is correspondence author.}

Except for the linguistic content that is recognized by ASR system, speech also conveys other important information about the speaker, such as voice,
emotion, gender, accent information etc.
As a result, recognition of such information is also widely studied~\cite{ivector,xvector,D.Povey-interspeech-2018-E2ESV,Sarma-2018-Interspeech-EID,J.Wang-2020-ICASSP-SER,Z.Wang-2017-ICASSP-GR,Najafian-SC-AID-2020} in speech research area.

Though speech related research is widely studied, few efforts are focused on combining different recognition tasks as mentioned to study jointly.
For example, one rarely combines speech recognition with gender recognition, or combines speech  recognition with accent recognition.
More often than not, they are studied separately. 
This contradicts our human speech recognition behavior. Intuitively, while human recognize speech content, they can also recognize the gender or accent, or even speaker emotion of the incoming speech simultaneously.

In this paper, we propose a joint approach to performing speech and accent recognition using DNN-HMM modeling framework, namely TDNNf~\cite{Povey-2018-interspeech-TDNNF}.
One of the advantages for the proposed method is that it can be easily deployed for real-time speech and accent recognition simultaneously.
Our RTF is $\sim$0.4 under normal 2.6 GHz CPU setting.
Meanwhile, we also submit another paper using Transformer-based End-to-end (E2E) approach~\cite{Jicheng-INTERSPEECH2021}. Though the E2E approach can yield better results taking advantage of out-domain data, the main limitation of the method is much
slower than TDNNf method, 


The paper is organized as follows. Section~\ref{sec:prior-work} is to review prior related work. Section~\ref{sec:proposed-joint-rec} presents the proposed joint speech and accent recognition method. Section~\ref{sec:data-spec} is our data specification. Section~\ref{sec:exp-setup} briefs the overall experimental setup and Section~\ref{sec:results} presents the results 
of the proposed methods. After that, we perform further analysis in Section~\ref{sec:analysis},
and draw conclusions in Section~\ref{sec:con}.

\section{Related work}
\label{sec:prior-work}
Recently, multilingual ASR has been drawing wide attention,
but most of the works are only concerned with bilingual case, such as 
code-switching ASR~\cite{Z.Zeng-2019-interspeech-CS,H.Xu-2018-interspeech-cs,P.Guo-2018-interspeech-cs}. In this paper, our ASR system is a multilingual ASR system, which is instead built with 8 accented English corpora. Besides, different from code-switching ASR, our ASR system is trained with utterances with only single accent, which is beneficial to utterance-based accent recognition.
For accent recognition, the goal is to classify speaker's accent according to their pronunciation peculiarity. One can perform the task using speaker  or gender identification method, such as i-vector~\cite{ivector}, x-vector~\cite{xvector} or End-to-End~(E2E)  neural network classifier based methods~\cite{D.Povey-interspeech-2018-E2ESV, attention-SR-is-2020}. 

Since E2E attention-based framework becomes popular, multilingual ASR is getting much simpler as
the number of the grapheme-letters/characters for each language is quite limited~(e.g. English  has only 26 letters).
One can simply merge the grapheme-letters/characters of different languages as recognition output. For instance, it is reported
a 50-language multilingual ASR system is built with E2E method in \cite{FACEBOOK-2020-Arxiv-MLASR}. However, it is only focused on speech recognition task. 
Also using E2E framework, 
\cite{Multi-dialect-ICASSP-2018} proposed a method to improve multi-dialect speech recognition performance by using dialect information as an external input vector in the LAS ASR model.
For joint systems,~\cite{S.W-ASRU-LID-2017} proposed a joint language identification and speech recognition method on 10 different languages. Its ASR results are impressive but far from the state-of-the-art on individual languages. Besides, its language identification results are not compared with conventional i-vector, x-vector, or other E2E methods. 
Furthermore, \cite{H.WX-JointTraining-IS2020} implemented the same structure as \cite{S.W-ASRU-LID-2017} and investigated its performance on many more languages. 


Different from prior work, we employ DNN-HMM to realize a joint speech and accent recognition system. The framework is only a multilingual ASR system, by appending grapheme with different accent identifier~(we are using position-dependent grapheme lexicon~\cite{FACEBOOK-2019-ASRU-hybrid}). To be thorough, we not only compare our ASR performance with conventional monolingual DNN-HMM ASR system, 
we also compare our accent recognition performance with x-vector-based classifier. Besides, we also conduct a series of analysis on the proposed methods under different scenarios.

\section{Joint speech and accent recognition}
\label{sec:proposed-joint-rec}
Technically,  we can think of each accent as different language, hence given a normal English word,  we make the word and its phone labels  different for each individual accent. We then merge the accented data  to train a joint multilingual ASR system.
From this perspective, not only can such an ASR system perform speech recognition, it can also 
recognize the accent.

During training, assuming word ``\texttt{hello}" is from \texttt{British} accent, our ASR lexicon should have an item like ``\texttt{hello\_BRT  h\_BRT\_WB e\_BRT l\_BRT l\_BRT o\_BRT\_WB}", realizing a British accent-based ``\texttt{hello}", that is represented as ``\texttt{hello\_BRT}" to differentiate with ``\texttt{hello}" from other accents. Here
the ``\texttt{\_BRT}" is the accent identifier, while ``\texttt{\_WB}" is word boundary grapheme identifier as recommended in ~\cite{FACEBOOK-2019-ASRU-hybrid}.

We have two methods to conduct joint recognition during decoding. 
One is to assume our ASR system as a monolingual system, but 
each word has different pronunciations, denoted as  ``\textbf{Mono-joint}''. Let us use word ``\texttt{hello}" as example again, and one of its pronunciations in decoding lexicon would be ``\texttt{hello  h\_BRT\_WB e\_BRT l\_BRT l\_BRT o\_BRT\_WB}". 
By this means, we demonstrate word-based accent encoding with the help of accent identifiers at phone level.
When recognizing speech, we output phone, as well as word sequences. By simply counting the accent identifiers for phone/word, we get the accent of
the recognized utterance. One of the drawbacks of the method lies in it cannot encode the accent on utterance level as each word in the utterance can be recognized as different accent.
Therefore, we propose another method, that is, not only is the phone label different, the word label is also different, here we name it as~``\textbf{Multi-joint}''. In other words, the decoding lexicon is the same with the training lexicon, and the ASR is a multilingual ASR system. The benefit of this method is  that
it can approximately realize  utterance-based accent encoding
as the language model 
(LM) has no direct cross-accent n-gram ever happened.

\section{Data Specification}\label{sec:data-spec}
\begin{figure} [ht]
    \centering
     \includegraphics[width=8cm]{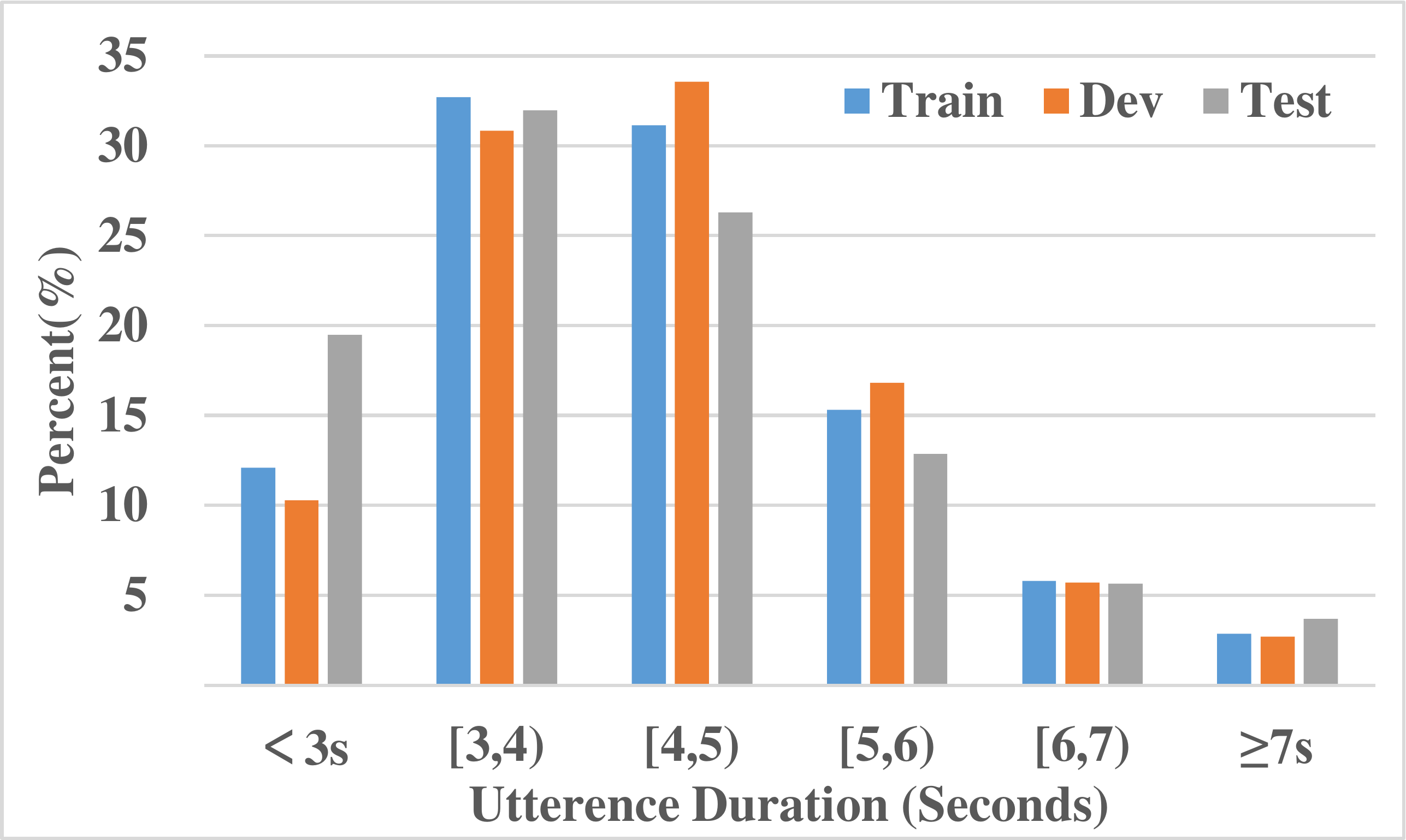}
    \caption{Utterance length(s) distribution for accented English data sets that are \texttt{train}, \texttt{dev} and \texttt{test} respectively}
    \label{fig:data-dist}
\end{figure}

The experimental data is from an accented English speech recognition workshop\cite{accented-english-icassp-2021}, sponsored by DataTang Company in China\footnote{https://www.datatang.com/INTERSPEECH2020}.
There are 2 challenge tracks, one is for accented English speech recognition, and the other is for accent recognition. The data contains two parts.
One is in-domain accented English data, released by DataTang, including \texttt{train}, \texttt{dev} and \texttt{test} sets specified by the organizer.
There are 8 accented English data sets among \texttt{train} and \texttt{dev} sets, each with $\sim$20 hours. The accents are American, British, Chinese, Indian, Japanese, Korean, Portuguese and Russian respectively. There are two extra accented English data sets on \texttt{test} set, which are Spanish and Canadian~(they are ignored when we perform accent recognition).
Table~\ref{tab:data-spec} reports the details of data specification. For clarification, Figure~\ref{fig:data-dist} also plots the utterance length distribution for the three data sets respectively.
The other is out-domain data which is 960 hours of Librispeech\footnote{http://www.openslr.org/12/}.

All data are read speech. As will be shown in Section~\ref{sec:results}, the WER is quite low for the ASR track. However, as is shown in Figure~\ref{fig:data-dist}, majority of utterance length is less than 6 seconds~(Table~\ref{tab:data-spec} also reports that the average utterance length is $\sim$4s) for the three data sets. Therefore, one can imagine the accent recognition is rather challenging for such short utterances. 


\begin{table}[htbp]
 \centering
 \caption{Data specification for the accented English data sets including \texttt{train}, \texttt{dev} and \texttt{test} respectively}
 \label{tab:data-spec}
\begin{tabular}{c|c|c|c}
 \hline
    & Train & Dev & Test \\
   \hline
   Total utts & 124K & 12k & 18k\\
Length (Hrs) & 148.51 & 14.50 & 20.95\\
Ave. word (per utt.) & 9.72 & 9.66 & 9.00 \\
Ave. second (per utt.) & 4.29 & 4.35 & 4.15 \\
    \hline
\end{tabular}
\end{table}

\begin{table*}[htb]
 \centering
 \caption{Speech WER~(\%) and Accent Recognition ACC~(\%) results of the proposed joint recognition methods on \texttt{dev} set.
 }
 \label{tab:Baseline_Comparison}
 \begin{tabular}{c | c c c c c c c c c c}
    \hline
    \hline
     & Method & AVE (WER\%) & US & UK & CHN & IND & JAP & KOR & POR & RUS \\
     \hline
      \multirow{3}*{\makecell{Speech \\ Recognition}} &  \tl {TDNNf} & 7.68 & 8.71 & 8.87 & 11.08 & 8.67 & 6.08 & 6.87 & 6.45 & 4.47 \\
     
     & \tl {Mono-joint} & 8.66 & 9.46 & 10.31 & 12.67 & 9.02 & 6.98 & 7.28 & 7.43 & 5.36 \\ 
     & \tl {Multi-joint} & 8.63 & 9.29 & 9.71 & 12.71 & 9.16 & 6.68 & 7.36 & 7.51 & 5.30 \\ 
     \hline
    &  & AVE (ACC\%) \\
    \hline
   \multirow{3}*{\makecell{Accent \\ Recognition}} & \tl {x-vector} & 64.86 & 55.05 & 88.24 & 55.44 & 75.25 & 55.28 & 56.04 & 79.58 & 52.91\\ 
     & \tl {Mono-joint} & 61.70 & 42.64 & 88.43 & 63.76 & 86.82 & 42.54 & 56.65 & 58.97 & 54.95 \\ 
     & \tl {Multi-joint} & 63.19 & 44.53 & 88.48 & 66.31 & 87.66 & 43.88 & 58.71 & 61.51 & 55.63 \\ 
     \hline
     \hline
 \end{tabular}
\end{table*}

\section{Experimental Setup}
\label{sec:exp-setup}
All the experiments are conducted with Kaldi.\footnote{https://github.com/kaldi-asr/kaldi}
For DNN-HMM ASR framework, the acoustic models are trained with Lattice-free Maximum Mutual Information~(LF-MMI) criterion~\cite{Povey-2016-interspeech-LFMMI}  over the Factorized Time Delay Neural Network~(TDNNf)~\cite{Povey-2018-interspeech-TDNNF}.
The TDNNf is made up of 15 layers, and each layer is decomposed as 1536x512x1536, where 512 is the dimension of the bottleneck layer. 
To train the TDNNf, we also perform two kinds of data augmentation~(DA) methods, one is speed perturbation~(sp) (x3)~\cite{ko-2015-speed_perturb}, and the other is to add four kinds of noise, such as white noise, music noise, babble noise, as well as reverberant noise (x4).
For baseline system, the lexicon
is phonetic lexicon with $\sim$21k word vocabulary.
For the multilingual ASR, the lexicon is position-dependent grapheme lexicon but with as many as eight times of the vocabulary for either two methods as mentioned.  From our off-line experiments, the grapheme lexicon yields slightly worse results, compared with phonetic lexicon. We stick to use it because we want our TDNNf system to have the same advantage with E2E ASR system, that is, free of lexicon modeling, which is especially useful for multilingual ASR tasks. For the two methods, we build two sets of tri-gram language models (LMs) for decoding respectively.
One is monolingual LM, and the other is a multilingual LM, labelling each word with different accent identifiers.

Additionally, we 
also perform x-vector experiment for
accent recognition as a contrast, while we choose logistic regression as the classifier. We find it consistently yields better results compared with PLDA~\cite{wang-2009-plda} 
method in our case.
 To train the x-vector extractor,
 we follow the configuration of~\cite{xvector}, and the resulting x-vector dimension is 512.

%
%

  \section{Results}\label{sec:results}
Table~\ref{tab:Baseline_Comparison} reports the speech recognition and accent recognition  results of the proposed methods, in contrast with conventional TDNNf, and x-vector systems respectively.

From  Table~\ref{tab:Baseline_Comparison},
we observe the proposed methods achieve worse ASR results compared with the baseline TDNNf system. This is understandable, since for the ``Mono-joint" method, we have 8-times of  pronunciations than the baseline system, yielding more confusion to the lexicon. Besides, the pronunciation difference for each specific accent is not fully considered. On the other hand, the TDNNf output of the ``Multi-joint" method has 8-times of senones to handle in theory, compared with the baseline TDNNf system. Such a big senone-number output is difficult to obtain by decision-tree-based clustering method, and it is also hard to learn.

For the accent recognition results in Table~\ref{tab:Baseline_Comparison}, the proposed methods also yield worse accent recognition results, compared with x-vector method. However, the advantage of the proposed method lies in two aspects. First to perform accent recognition, it is easier for the proposed method to exploit out-domain data (\texttt{Librispeech} here) without accent label, compared with x-vector method. Such an advantage will be shown in Section~\ref{ssec:transfer_Learning}. Secondly, it can perform both speech and accent recognition tasks simultaneously. 

Besides, from Table~\ref{tab:Baseline_Comparison}, we observe the best WERs are from Russian (RUS) and Japanese (JAP) accented speech respectively, while the worst of all is from Chinese (CHN) accented speech, which is followed by British (UK) accented speech. For the accent recognition, the best accuracy is from British accent (UK), and after that it is Indian accent (IND), while the worst is American accent (US) which is followed by the Japanese accent (JAP). One thing worth a note is that the proposed method outperforms the x-vector method with a big margin for the Chinese (CHN) and Indian (IND) accented speech respectively, while it performs much worse on the American (US), Japanese (JAP) and Portuguese (POR) accented speech.
Furthermore,
comparing speech recognition with accent recognition results in Table~\ref{tab:Baseline_Comparison},
we note that better WER does not necessarily mean better accent recognition accuracy. For instance, for the Korean (KOR), Japanese (JAP), and Russian (RUS) accented speech, the ASR results are significantly better than the ones of remaining accents, their accent recognition results are much worse.

Finally, Table~\ref{tab:Baseline_Comparison} shows that the proposed ``Multi-joint" method outperforms the ``Mono-joint" method for both speech and accent recognition results respectively. 
This suggests utterance-based accent encoding is superior than the word-based accent encoding at least for accent recognition.

\begin{table}
    \centering
    \caption{Performance report with different TDNNf outputs (tied-states/senones), dimension of the final bottleneck layer is fixed at 512.}  
    \begin{tabular}{c|c c|c c}
    \hline
      \multirow{2}*{\#PDF} & \multicolumn{2}{c|}{Mono-joint} & \multicolumn{2}{c}{Multi-joint} \\
       & WER\% & ACC\% & WER\% & ACC\% \\
        \hline
        4.3k & 8.78 & 60.27 & 8.78 & 61.68 \\
        \hline
        \textbf{8.8k} & \textbf{8.66} & \textbf{61.70} & \textbf{8.63} & \textbf{63.19} \\
        \hline
        14.6k & 8.76 & 60.97 & 8.84 & 61.99 \\
        \hline
        21.8k & 8.76 & 60.97 & 8.80 & 62.28 \\
        \hline
    \end{tabular}

    \label{tab:Senone}
\end{table}

\section{Analysis}
\label{sec:analysis}

\begin{table*}[htp]
    \centering
    \caption{Cross-Accent Recognition Confusion on \texttt{dev} set(\%)} 
    \label{tab:confusion}
    \begin{tabular}{c}
        \begin{minipage}[t]{0.81\linewidth}
            \includegraphics[width=1\linewidth]{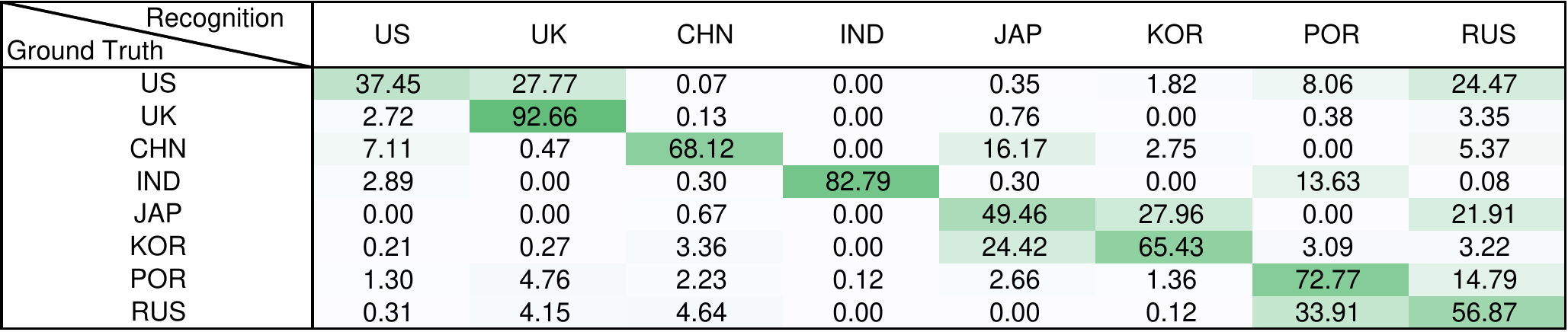}
        \end{minipage}
    \end{tabular}
\end{table*}

\subsection{Joint accent acoustic models}
\label{sub:pdf-number}
For the above experiments, we merge 8 accented English data to train the joint TDNNf acoustic models, and each accent has its own phone set.  The TDNNf models have $\sim$8.8k outputs (tied-states/senones) in total. As a result, we only have about 1k outputs for each accent on average. We are curious that if bigger outputs yield better results. Table~\ref{tab:Senone} reports our experimental results. From Table~\ref{tab:Senone}, bigger outputs of TDNNf don't bring performance improvement on both tasks. We also report the results of a smaller output of ~4.3k where we only have less than 6 hundred output on average for each accent, and get a slightly worse performance compared with those of larger scales of outputs. Our conclusion is that TDNNf is not sensitive to the tied-state number~(perhaps once it is satisfied with a minimal tied-state number).
This even suggests we can use TDNNf to train the joint multilingual models with more languages in future.

 
\subsection{Transfer Learning}
\label{ssec:transfer_Learning}
\begin{figure}[htbp]
    \centering 
    \begin{tabular}{cc}
        \begin{minipage}[t]{0.49\linewidth}
            \includegraphics[width=1\linewidth]{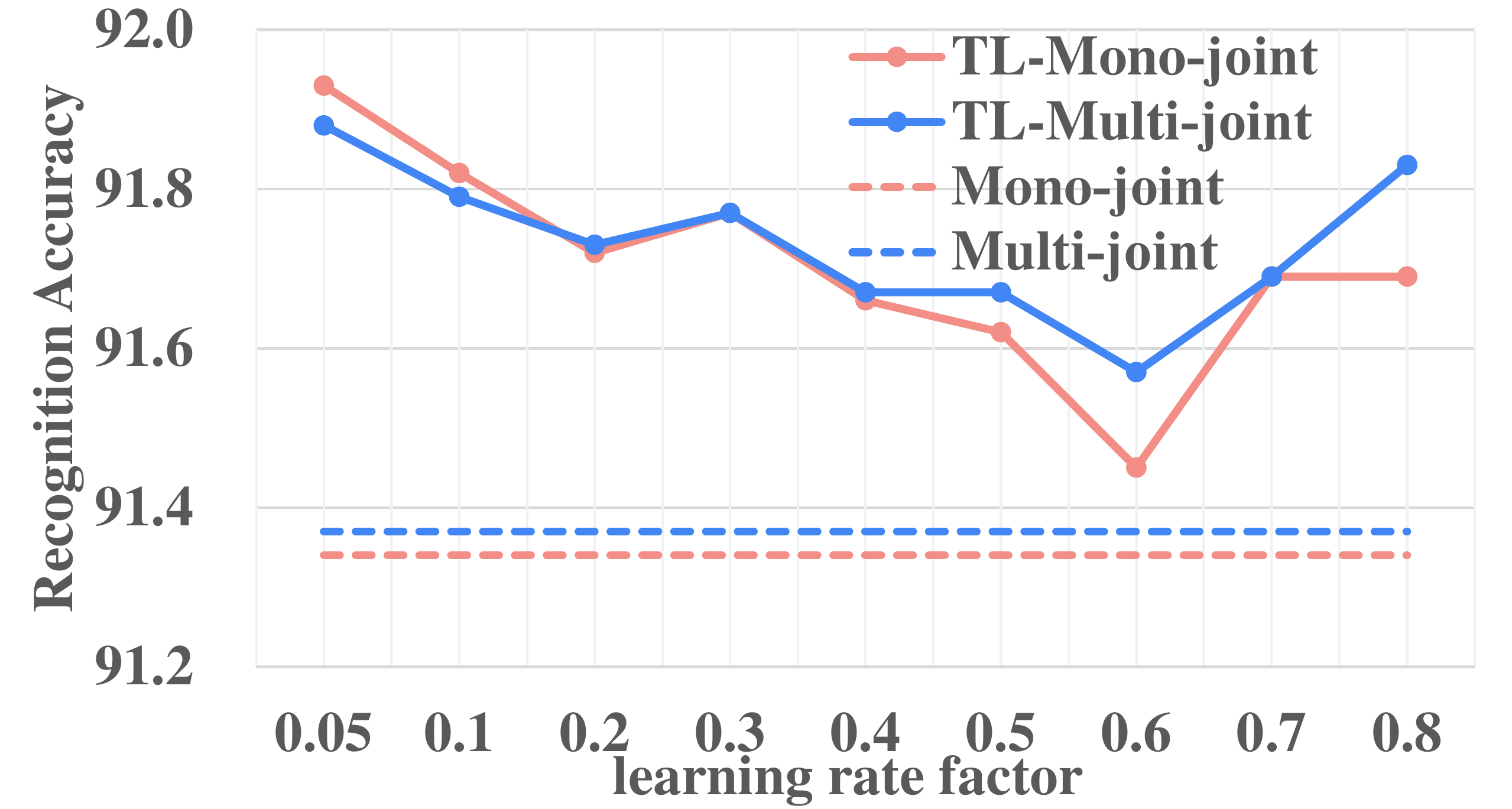}
            \centerline{(a) Speech Recognition}\medskip
        \end{minipage}
        \begin{minipage}[t]{0.505\linewidth}
            \includegraphics[width=1\linewidth]{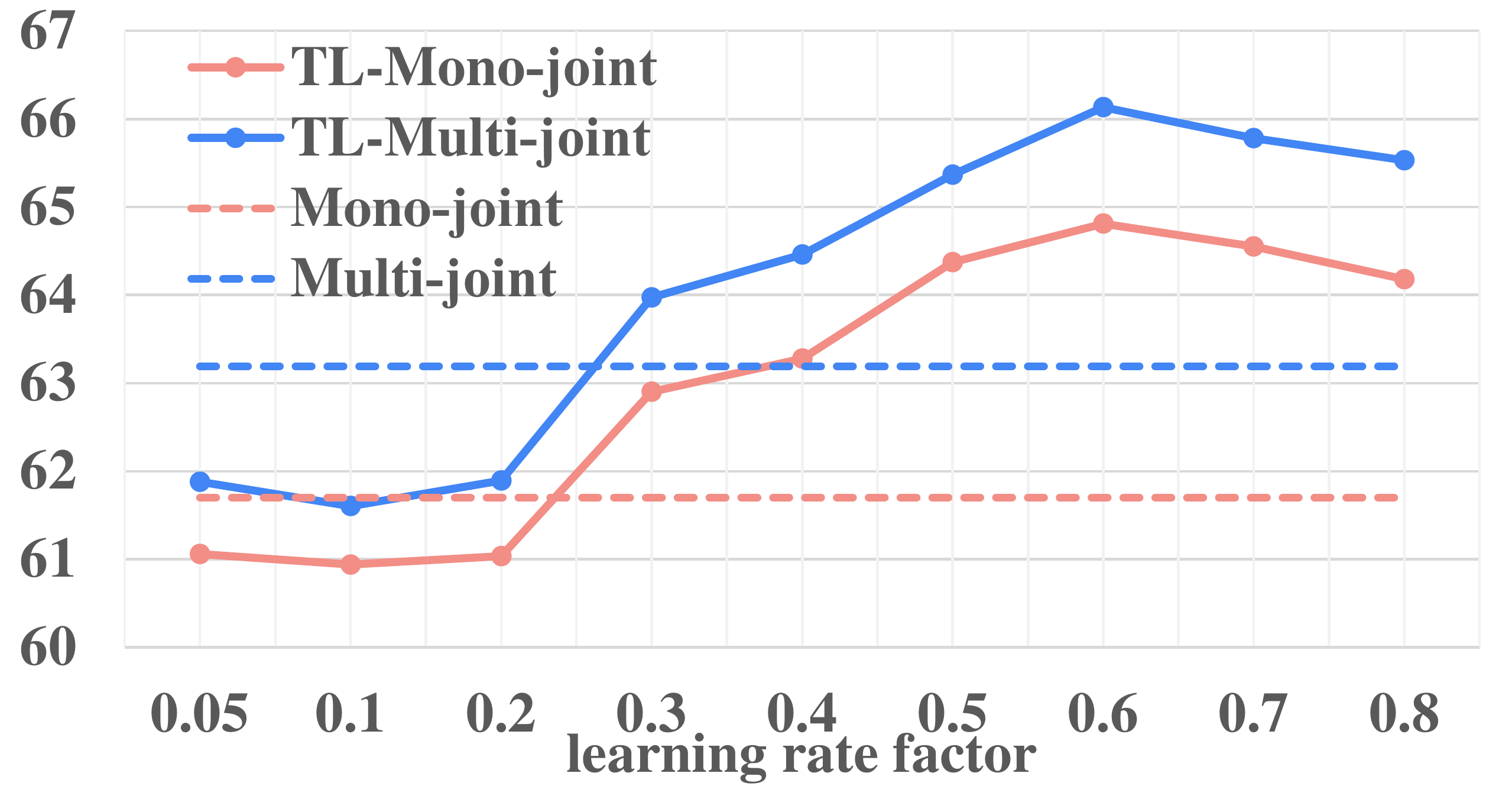}
            \centerline{(b) Accent Recognition}\medskip
        \end{minipage}
    \end{tabular}
    
    \caption{TL performance versus learning rate factor}
    \label{fig:Transfer-learning}
\end{figure}

To employ out-domain data, we adopt transfer learning~(TL) method~\cite{P-2017-ASRU-transfer-learning,Transfer-learning-WangDong-APSIPA2015, Transfer-learning-Huang-Neurocomputing}, with \texttt{Librispeech} as the out-domain data. 
We start with a TDNNf model trained with overall out-domain and in-domain data.
We then fine-tune the model with only the in-domain accented data. Figure~\ref{fig:Transfer-learning} plots the curves of the recognition accuracy versus learning rate factor, with 3 epochs. From Figure~\ref{fig:Transfer-learning}, we observe consistent improvements on both tasks with bigger learning rate factor ($\geq$0.5).

Table~\ref{tab:tl-rnnlm-rescore} reports the results of transfer learning, and RNNLM rescoring. From Table~\ref{tab:tl-rnnlm-rescore}, we see that transfer learning is effective on both speech and accent recognition performance improvement,
even outperforms x-vector (64.86\%) on accent recognition.
However, RNNLM rescoring only improves ASR WERs, while it yields no accuracy improvement on 
AR task at all.
These suggest AR accuracy is closely related with the performance of the underlying acoustic model (AM), but less affected with LM. For AM, different acoustic senones are trained with different accented speech data, as a result it has capability to differentiate accents.  However, for  the LMs (RNNLM included), they are  trained with overall transcripts, and consequently have no accent discriminative capability  on either word or utterance level.


%




\begin{table}[htbp]
    \centering
        \caption{Recognition results with TL(lr=0.8) and RNNLM Rescoring for \texttt{dev} and \texttt{test} sets}
    \begin{tabular}{c|c c|c c}
        \hline
        \multirow{2}*{Method} & \multicolumn{2}{c|}{WER(\%)} & \multicolumn{2}{c}{ACC(\%)} \\
         & dev & test & dev & test \\ 
        \hline
        \tl {Mono-joint} & 8.66 & 10.31 & 61.70 & 61.01 \\
        \tl {\quad +Transfer Learning} & 8.31 & 9.83 & 64.18 & 63.70  \\
        \tl {\quad +RNNLM Rescoring} & 6.93 & 8.28 & 64.29 & 62.66  \\
        \hline
        \tl {Multi-joint} & 8.63 & 10.35 & 63.19 & 63.75 \\
        \tl {\quad +Transfer Learning} & 8.17 & 9.74 & 65.53 & 66.68  \\
        \tl {\quad +RNNLM Rescoring} & 7.05 & 8.40 & 65.35 & 66.40\\
        \hline
    \end{tabular}
    \label{tab:tl-rnnlm-rescore}
\end{table}

\subsection{Cross-accent recognition confusion}
\label{ssec:Accent-recognition Ratio}
We are also  interested in how each specific accent is mistakenly recognized as  another accent, that is, cross-accent recognition confusion pattern.
Table~\ref{tab:confusion} shows overall accent recognition confusion results. There are several patterns worth our notice.
First, for British and Indian accents, there are minor in confusion with other accents.
The most confusion counterparts for British and Indian accents are Russian and Portuguese ones, with 3.35\% and 13.63\% respectively.
Second, Japanese and Korean accents, as well as Portuguese and Russian accents are mutually confused heavily. For instance, Japanese accent has 27.96\% been miss recognized with Korean accents which in return has 24.42\% been miss recognized with the Japanese accent.
Finally, American accent is widely ``overlapped" with other accents, particularly 27.77\% with British, and followed by Russian (24.47\%), and then Portuguese (8.06\%) accents respectively.


\subsection{Accented-word characteristics}
\label{ssec:Characteristics}
Except for accent confusion, we  are also curious about if some words have stronger accented characteristics than others. In other words, if a  word appears in some accent and it is correctly recognized in terms of  accent recognition in majority of times, we think such a word has stronger accented characteristics.
Table~\ref{tab:exemplar-words} reveals some exemplar words in terms of highest and lowest accent recognition accuracy.  We can see from Table~\ref{tab:exemplar-words}, longer words or words with multiple syllables generally yield better accent recognition results. This seems to be agreed with our intuition.




\begin{table}[htbp]
    \centering
     \caption{Exemplar words with highest and lowest accent recognition accuracy respectively}
    \begin{tabular}{c|c|c}
        \hline
       Top 10 & Exemplar words& \makecell{Ave. Length} \\
         \hline
        \makecell{most accurate} & \makecell[l]{ \textcolor{red}{temperature, strong, feature,} \\ \textcolor{red}{students, adapted, voyage,} \\ \textcolor{red}{branches, value, plans, blind}} & 7.3\\ 
                \hline
        \makecell{least accurate } & 
            \makecell[l]{\textcolor{blue}{fake, key, finish }\\
                      \textcolor{blue}{minute, England, ocean,} \\
                    \textcolor{blue}{orders, sick, either, wash}} & 5.4 \\ 
                \hline
    \end{tabular}
    \label{tab:exemplar-words}
\end{table}

\section{Conclusions}\label{sec:con}
In this paper, we proposed a joint multilingual approach to realizing speech and accent recognition simultaneously using the DNN-HMM framework. On an 8-accent English speech recognition data set, we demonstrated the effectiveness of the proposed methods on both speech recognition and accent recognition. In addition, we adopted transfer learning to fully exploit out-domain data to boost the performance. We found that transfer learning is beneficial to the performance improvement for both tasks, particularly for the accent recognition task. We also tried RNNLM rescoring method, and achieved better WER as expected however no accent recognition improvement. This suggests accent recognition is more dependent on AM and less on LM for our data set.
Besides, we also analysed cross-accent recognition confusion, as well as accented-word characteristics preliminarily, which are yet to be  further studied in future.

\vfill\pagebreak
\clearpage
\bibliographystyle{IEEEtran}
\bibliography{refs}

\end{document}